# Statistical Postprocessing for Weather Forecasts – Review, Challenges and Avenues in a Big Data World

Stéphane Vannitsem[11,14], John Bjørnar Bremnes[10], Jonathan Demaeyer[11,14], Gavin R. Evans[8], Jonathan Flowerdew[8], Stephan Hemri[4], Sebastian Lerch[6], Nigel Roberts[9], Susanne Theis[2], Aitor Atencia[13], Zied Ben Bouallègue[3], Jonas Bhend[4], Markus Dabernig[13], Lesley De Cruz[11], Leila Hieta[5], Olivier Mestre[7], Lionel Moret[4], Iris Odak Plenković[1], Maurice Schmeits[12], Maxime Taillardat[7], Joris Van den Bergh[11], Bert Van Schaeybroeck[11], Kirien Whan[12], Jussi Ylhaisi[5]

1. Croatian Meteorological and Hydrological Service, Borongajska cesta 83d/1, 10000 Zagreb, Croatia
2. Deutscher Wetterdienst, Frankfurter Straße 135, 63067 Offenbach, Germany
3. European Center for Medium-range Weather Forecasts, Shinfield Park, Reading, United Kingdom
4. Federal Office of Meteorology and Climatology, MeteoSwiss, Operation Center 1, 8058 Zürich-Flughafen, Switzerland
5. Finnish Meteorological Institute, Erik Palménin aukio 1, FI-00560 Helsinki, Finland
6. Karlsruhe Institute of Technology, Institut für Stochastik, Englerstr. 2, D-76131 Karlsruhe, Germany
7. Météo-France, CNRM - UMR 3589, Toulouse, France
8. Met Office, Exeter, United Kingdom
9. MetOffice@Reading, Met Office, United Kingdom
10. Norwegian Meteorological Institute, Box 43 Blindern, Oslo, Norway
11. Royal Meteorological Institute of Belgium, Avenue Circulaire, 3, 1180 Brussels, Belgium
12. Royal Netherlands Meteorological Institute (KNMI), P.O. Box 201, 3730 AE De Bilt, The Netherlands
13. Zentralanstalt für Meteorologie und Geodynamik, Hohe Warte 38, 1190 Wien, Austria
14. European Meteorological Network (EUMETNET), Avenue Circulaire, 3, 1180 Brussels, Belgium

Corresponding author: Stéphane Vannitsem, Stephane.Vannitsem@meteo.be

**Capsule Summary**: State-of-the-Art statistical postprocessing techniques for ensemble forecasts are reviewed, together with the challenges posed by a demand for timely, high-resolution and reliable probabilistic information. Possible research avenues are also discussed.

**Summary**

Statistical postprocessing techniques are nowadays key components of the forecasting suites in many National Meteorological Services (NMS), with for most of them, the objective of correcting the impact of different types of errors on the forecasts. The final aim is to provide optimal, automated, seamless forecasts for end users. Many techniques are now flourishing in the statistical, meteorological, climatological, hydrological, and engineering communities. The methods range in complexity from simple bias corrections to very sophisticated distribution-adjusting techniques that incorporate correlations among the prognostic variables.

The paper is an attempt to summarize the main activities going on this area from theoretical developments to operational applications, with a focus on the current challenges and potential avenues in the field. Among these challenges is the shift in NMS towards running ensemble Numerical Weather Prediction (NWP) systems at the kilometer scale that produce very large datasets and require high-density high-quality observations; the necessity to preserve space time correlation of high-dimensional corrected fields; the need to reduce the impact of model changes affecting the parameters of the corrections; the necessity for techniques to merge different types of forecasts and ensembles with different behaviors; and finally the ability to transfer research on statistical postprocessing to operations. Potential new avenues will also be discussed.

# 1. Introduction

Errors from multiple sources have a detrimental effect on the skill of weather forecasts. One of the primary sources of errors is associated with the construction of an initial condition for numerical weather forecasting systems. These errors grow rapidly during the course of the forecasts until they reach a level beyond which the forecasts do not display any useful skill. This property is known as the sensitivity to initial conditions, see e.g. the review of Vannitsem (2018). Two other important categories of errors that reduce forecast skill are the boundary-condition errors (e.g. Collins and Allen, 2002; Nicolis, 2007) and the model structural errors. Model structural errors include a missing or poor representation of sub-grid dynamical and physical processes and inaccuracies associated with the numerical scheme (Lorenz, 1982; Nicolis et al, 2009). All these Numerical Weather Prediction (NWP) model deficiencies affect the course of the forecasts by inducing what are often called systematic and random errors.

Since the early 1950s, numerical weather systems have been developed with an ever increasing complexity (Lynch, 2007). In conjunction, the quality of the forecasts has been constantly improving, with current forecasts deemed useful even beyond 15 days (Buizza and Leutbecher, 2015; Bauer et al, 2015). This success can be attributed to both improvements in the quality of the initial conditions of the numerical prediction models coming from improved data assimilation, and improvements in the model representation of physical and dynamical processes; or in other words a reduction of both initial condition and model errors. Nowadays, many operational forecasting centers are running both single high-resolution deterministic forecasts (at km-scale grid spacing) and ensemble forecasts (often at km-scale too), the latter providing information about the probability of occurrence of specific atmospheric states (Toth and Kalnay, 1993; Molteni et al, 1996; Yoden 2007; Leutbecher and Palmer, 2008; Buizza et al, 2019; Frogner et al, 2019).

Despite these major developments, both deterministic and ensemble forecasts continue to display significant deficiencies associated with the presence of model errors and an inappropriate distribution of initial-conditions. This results in systematic biases and inappropriate dispersion of ensemble forecasts requiring some sort of postprocessing in order to improve the forecast quality. Statistical postprocessing methods used for this purpose involve a wide range of correction techniques that can be appropriately developed for either deterministic or ensemble forecasts (Wilks, 2011; Vannitsem et al, 2018).

The first applications and operational implementations of statistical corrections were based on simple linear regression techniques using linear statistical relations deduced from observational data only, known as Perfect Prog, or built between the observations and predictors generated by the weather forecasting models, known as Model Output Statistics (Klein et al, 1959; Glahn and Lowry, 1972). These approaches were successful, attracting much interest from the meteorological community with many applications to a wide range of model variables and extensions to other types of regression functions like logistic regressions or neural networks (e.g. Lemcke and Kruizinga, 1988; Marzban, 2003). Nowadays there is a bloom of techniques, in particular, for ensemble forecasts with the purpose of producing probabilistic information with a more accurate representation of forecast uncertainty (Gneiting et al, 2007).

Since postprocessing aims to improve the quality and usefulness of the forecasts, an important aspect is the choice of measures used to assess that quality. Three key attributes of ensemble forecasts are usually sought: First, the forecasts should be as close as possible to the truth or the observations (the proxy for the truth); second, the forecast should respect the climatology or the frequency with which different thresholds are exceeded, a climatological reliability; and third, the observation should be statistically indistinguishable from the forecasts produced by the model. The first property is related to the resolution/sharpness of the forecasts and the two latter to the reliability. Statistical calibration provides a natural way to improve reliability, essential for rational decision making, sometimes at the expenses of resolution/sharpness. Verification can help to identify the key systematic errors which the postprocessing should be designed to address, check its success in correcting them, and the impact on wider performance measures. A variety of scoring rules and diagnostic tools exists for this purpose (e.g. Richardson, 2000; Wilks 2006; Schuhen and Thorarinsdottir, 2018). For ensemble forecasts, Brier skill scores, Rank Probability Skill Scores, Rank histograms or Spread-Skill relationships, are popular measures.

Statistical correction techniques for both deterministic and ensemble forecasts should nowadays be an integral part of any operational forecasting system. As illustrated in Hemri et al (2014), whatever the degree of sophistication of the model under consideration, the statistical postprocessing approach still provides additional corrections that will benefit the end user. This last consideration should make research into this area and operational implementation key priorities of National Meteorological Services (NMS).

The goal of this paper is to review the current research developments and operational implementations taking place worldwide and particularly in Europe, together with the future prospects and challenges in the area of statistical postprocessing. A key challenge is the exponential growth of data that are available from both the model forecasts and the observations, accompanied by an ever increasing need for very localized, yet seamless, forecast information.

Section 2 discusses state-of-the-art approaches for statistical postprocessing of ensemble forecasts. The impact of using statistical postprocessing on the physical coherence of the dynamics is addressed in Section 3. Section 4 covers the problem of the frequent model changes that could affect the quality of the statistical correction techniques. In section 5, the use of blending techniques for correcting the forecasts and providing seamless probabilistic information is reviewed. Section 6 evaluates the potential implementation difficulties of the statistical correction techniques. Finally, future prospects and challenges are discussed in Section 7.

**2. State-of-the-art statistical postprocessing methods**

From a statistical perspective, most postprocessing methods can be categorized into two groups - those that assume the predictive distribution belongs to a class of known probability distributions (parametric approaches) and those that do not (nonparametric approaches). Recent developments for these two categories will be addressed separately, and then some available tools for both are listed. Some key methodological challenges are briefly discussed at the end of the section.

   a. *Development of parametric approaches*

Parametric postprocessing approaches specify a parametric model for the forecast distribution by selecting a suitable family of probability distributions depending on the variable of interest (Gneiting et al, 2005; Raftery et al, 2005). The parameters of the forecast distributions are then linked to the predictors from the NWP system via regression equations to correct systematic errors. The regression coefficients are estimated by optimizing suitable loss functions for distribution forecasts such as the continuous ranked probability score (CRPS; Gneiting and Raftery, 2007). Such constructions lead to models that are straightforward to fit and are widely used at NMS. Summary statistics of the ensemble predictions of the variable of interest are often used as sole covariates. However, usually many more potential predictors (including predictions of other variables as well as geographical or temporal information) are available and might provide great benefit, but specifying their functional relations to the distribution parameters is challenging. One promising approach is gradient boosting for distributional regression (Messner et. al., 2016) which selects the most important predictor variables during parameter estimation by iteratively updating the regression coefficient of the predictor that improves the current model fit the most.

Several variants where the forecast distribution remains parametric, but fewer assumptions concerning the functional relation between predictors and distribution parameters are required, have been proposed recently. Lang et al. (2019) and Simon et al. (2019) use generalized additive models where the parameters of a forecast distribution from a wide range of distributional families with parameterizations corresponding to location, scale, and shape are flexibly modelled as additive functions of the predictors (Rigby and Stasinopoulos, 2005). The functional dependencies between distribution parameters and predictors are usually prescribed separately for each parameter. Schlosser et al. (2019) extend this framework by using regression trees and random forests to recursively partition the predictor space based on maximum likelihood estimation, resulting in separate distribution models for each partition. Rasp and Lerch (2018) propose the use of neural networks to link distribution parameters to predictors. Since neural networks are very flexible functions composed of a sequence of nonlinear transformations, they facilitate modeling exceedingly flexible relations jointly for all distribution parameters. By estimating the neural network weights/coefficients using optimization based on stochastic gradient descent algorithms, highly complex models can be fitted.

Despite these powerful variants the need to select a suitable parametric family to describe the distribution of the target variable remains a limitation for parametric postprocessing methods and often necessitates elaborate fine-tuning (Gebetsberger et al., 2017) or complex mixture models (Baran and Lerch, 2016) to achieve well-calibrated forecast distributions. A flexible alternative to the full-distribution adjusted parametric methods just discussed above is the kernel dressing method, consisting of replacing individual ensemble members by kernel functions (Roulston and Smith, 2003; Bröcker and Smith, 2008). This approach can accommodate any type of ensemble continuous distribution, but is also a parametric approach since the parameters of the kernel need to be fitted.

### b. Development of nonparametric approaches

An alternative is to use nonparametric postprocessing methods that avoid distributional assumptions by constructing nonparametric approximations of the forecast distribution. Bremnes (2004) proposes the use of quantile regression methods in which only a selected set of quantiles of the predictive distribution is considered. The quantiles are traditionally estimated separately which can lead to invalid crossing quantiles. However, by adding constraints to the estimation or by simply reordering the quantiles at the end, this problem can be circumvented. Quantile regression methods have recently been extended in several directions. Wahl (2015) proposes a penalization method in a Bayesian context where regularization and predictor selection are performed simultaneously. Ben Bouallègue (2017) suggests a similar penalized estimation approach that has proved useful in situations with many predictors. In Bremnes (2019) constrained splines are applied to make use of information from all ensemble members. Standard quantile regression methods are not suitable for extreme quantile levels and were the motivation of Velthoen et al. (2019) for developing an estimator for extreme levels based on local quantile regressions. Möller et al. (2018) apply D-vine copula regression methods to model the joint distribution of the weather variable of interest and the predictors, and derive the forecast distribution from this.

Cannon (2018) and Bremnes (2020) shift the focus from a finite set of quantiles to complete quantile functions, thus providing nonparametric models of the entire forecast distribution. The former adds the quantile level as a monotone predictor in a neural network approach, while the latter assumes the quantile function to be a Bernstein polynomial and model the relation between its coefficients and the predictors by means of a neural network like Rasp and Lerch (2018). Similarly targeting the full predictive distribution, Veldkamp et al. (2020) apply a discretization to the target variable and nonparametrically model the predictive density by a histogram, with probabilities obtained as output of a convolutional neural network. Isotonic distributional regression, a nonparametric technique that takes advantage of partial order relations among the predictors and requires minimal implementation decisions only was recently proposed by Henzi et al. (2019).

Taillardat et al. (2016) propose a postprocessing model using quantile regression forests, a quantile regression method where predictive quantiles are computed based on random forests (Breiman, 2001; Meinshausen, 2006). Random forest methods have been used in a variety of postprocessing applications (Gagne et al., 2009; 2017; McGovern et al., 2017) and quantile regression forests have been applied to a wide range of weather variables (Taillardat et al., 2016; Zamo, 2016; van Straaten et al., 2018; Whan and Schmeits, 2018). Recent extensions include combinations with parametric models fitted to forest-based outputs to circumvent the restriction of predictive quantiles by the range of training observations and to provide better forecasts for extreme events (Taillardat et al., 2019), as well as quantile regression forests calibration based on forecast anomalies to improve predictions of cold and heat waves (Taillardat and Mestre, 2020). The now operational ecPoint methodology (Pillosu and Hewson, 2017; Hewson and Pillosu, 2020) utilizes a decision-tree method based on expert elicitation to derive, for the whole world, point-wise precipitation forecasts from grid-box forecasts by accounting for both the expected grid-scale NWP bias and the expected sub-grid variability, according to the diagnosed grid-box weather type.

Further, non-parametric postprocessing methods based on historical analogs can be viewed in a similar framework (Taillardat et al., 2016). Instead of obtaining sets of analogous historical cases by recursively partitioning the predictor space via random forests, analog-based methods (Hamill and Whitaker, 2006; Delle Monache et al., 2011; 2013) sequentially search for the most similar past cases for every new input vector of predictor values using a specifically tailored similarity measure, usually based on weighted combinations of several predictors. Forecast distributions are then constructed from the corresponding set of past observations. Analog-based methods have been applied for a variety of variables (Alessandrini et al., 2015, 2018; Nagarajan et al., 2015; Odak Plenković et al., 2018; 2020). As for all postprocessing methods, there is a clear benefit of increasingly large training datasets: in this case, the benefit is that closer analogs can be found. However, the computational cost of determining suitable analogs may become prohibitive for large datasets, particularly in the presence of many possible covariates. Several studies have proposed the use of analogs or related concepts to determine training datasets consisting of similar past forecast cases for parametric (Hamill et al., 2008; Junk et al., 2015; Lerch and Baran, 2017; Scheuerer and Hamill, 2019; Schlosser et al., 2019) and non-parametric (Bremnes, 2004; Hamill et al., 2015) methods. These approaches are motivated by the premise that customized training datasets consisting of similar past forecast cases may allow the use of simpler models that are better adapted to the current conditions, and are closely related to local modeling methods in statistics (see, e.g., Bottou and Vapnik, 1992; Loader, 1999).

Other non-parametric postprocessing methods use various direct transformations of ensembles, emerging from simple bias correction or variance inflation procedures (e.g., Johnson and Swinbank, 2009). Flowerdew (2014) directly maps forecast probability to observed event frequency for a series of thresholds, to make the forecasts reliable (and thus unbiased and correctly spread) whilst preserving the resolution. Each threshold is calibrated using the most local region consistent with achieving a specified sample size, and the algorithm only needs to see each piece of training data once, reducing storage requirements for large gridded datasets. Approaches acting simultaneously on each member in an ensemble have been suggested by Van Schaeybroeck and Vannitsem (2015) and others. In these, all parameters defining the transformation of the ensemble are estimated jointly by either optimizing the empirical CRPS or following maximum likelihood principles under further distributional assumptions. Some available tools are listed in the Appendix.

  *c. New Methodological challenges*

Most new methodologies are based around Machine Learning (ML) techniques. Although first attempts at using modern ML approaches for postprocessing have shown promising improvements over traditional approaches, a number of challenges remain. It is important that inherent structures in NWP forecasts (and in NWP errors) should be more fully exploited. For example, current approaches often rely on a limited set of ensemble statistics and covariates only, but modern ML approaches, in particular approaches involving neural networks, could be designed to target physical relationships between a large number of covariates more directly. These approaches can also be efficiently implemented on massively parallel supercomputers (e.g. Cervone et al, 2017). Furthermore, recent developments in deep learning, notably convolutional neural networks, have made it possible to use large gridded input data sets, allowing one to use more spatial information in statistical postprocessing, see Veldkamp et al. (2020) for a first application.

A second set of challenges relates to the interpretability of ML approaches. Whilst many methods are regarded as 'black boxes' various techniques can provide an understanding of what ML models have learned (see McGovern et al., 2019 for a recent review), for example which predictors are most important in the model and for a particular forecast. Many approaches report global variable importance (Taillardat et al., 2016, Rasp and Lerch, 2018, van Straaten et al., 2018; Whan and Schmeits, 2018) that can be used as a sanity check for the model and can give insights about relationships between the response and the set of predictors. An explanation of individual predictions is of interest to many users and can be achieved with methods such as Shapley values (Molnar, 2019). Due to their complexity, modern ML approaches, and deep neural networks in particular, can produce unexpected results at times.

## 3. Preserving space and time correlation

Weather forecasts typically include spatial, temporal and multi-variable information. Forecasters, and even the public, are nowadays accustomed to animated maps of multiple meteorological variables. Such visualizations reveal common patterns in space, time and between variables. These range from trivial relations (e.g. clustering of rainfall cells) to more complex relations (e.g. between radiation, temperature and humidity), and together are called the dependence structure. Given the assumption that a modern NWP model can reproduce the correct dependence structure, the question is whether postprocessing methods are able to do so as well and hence adjust variables in a consistent way. This is also important in the context of downstream applications such as hydrological ensemble forecasting systems and renewable energy applications (Cloke and Pappenberger, 2009; Pinson and Messner, 2018), for which space-time coherence is crucial for realistic scenarios. For weather variables with continuous probability distributions such as temperature or wind, the dependence structure is automatically preserved by the so-called member-by-member (MBMP) approaches (Doblas-Reyes et al., 2005; Johnson and Bowler, 2009; Flowerdew, 2014; Van Schaeybroeck and Vannitsem, 2015). In contrast, techniques that use predictive distributions require additional approaches to re-establish the dependence structure (Schefzik and Möller, 2018). One approach is based on parametric methods and makes use of specific multivariate predictive distributions that are suitable for low-dimensional settings or settings with specific intervariable relations (Pinson and Girard, 2012). Higher-dimensional situations can be adequately handled by nonparametric methods such as ensemble copula coupling (ECC, Schefzik et al., 2013) and the Schaake shuffle (SSH, Clark et al., 2004, Sperati et al., 2017). ECC is based on empirical copulas aimed at restoring the dependence structure of the forecast and is derived from the rank order of the members in the raw ensemble forecast, under a perfect model assumption, with exchangeable ensemble members. For SSH, on the other hand, the dependence structure is derived from historical observations instead. Finally, dependencies can also be taken into account through univariate postprocessing methods with location-specific model parameters. These methods include geostatistical model averaging (Kleiber et al., 2011) and spatially adaptive models that make use of anomalies (Scheuerer and König, 2014) or standardized anomalies as in standardized anomaly model output statistics (Dabernig et al., 2017; 2020). Figure 2 illustrates two specific routes of postprocessing that allows the preservation inter-variable dependences for high-dimensional problems.

Despite their simplicity and efficiency, multivariate approaches like ECC or SSH are prone to introduce physically unrealistic artifacts into the forecast scenarios. SSH is not flow-dependent and ECC is impaired by small spread and errors in the dependence structure of the raw ensemble (Schefzik and Möller, 2018). Flow-dependent variants of the SSH that select historical observations for the dependence template based on some similarity measures have been developed by Schefzik (2016), Scheuerer et al. (2017), and Scheuerer and Hamill (2018). Ben Bouallègue et al. (2016) propose dual ECC as an ECC variant which corrects for errors in the dependence template by also considering the auto-correlation of past forecast errors. Furthermore, if the dependence template, i.e. the raw ensemble for ECC, includes many equal values, e.g. zeros for precipitation, reordering of quantiles of the post-processed predictive distribution is not straightforward. Random reordering would lead to forecast scenarios with physically unrealistic jumps between spatially and temporally close locations. Resolving this issue may include methods by Scheuerer and Hamill (2018) to generate realistic forecast scenarios in the presence of ties or the modifications to ECC in order to smooth sharp jumps by Bellier et al. (2018).

The method of analogs already discussed in Section 2 is an approach which easily allows the introduction of spatial and temporal correlations in the output. For instance, comparing the entire fields or objects in the analog-search instead of independent point-by-point events and then associating the analog to an analysis provides a post-processed solution possessing the spatial (and temporal) correlation of the analysis (e.g. Hamill and Whitaker, 2006; Frediani et al, 2017).

**4. Coping with model changes**

Many of the methods presented in the preceding sections rely on the availability of large archives of past forecast and observation data, and it is usually assumed that the error characteristics do not change substantially over time. However, this assumption is often threatened by inhomogeneous model and observation data brought about by changes in the observation systems and upgrades to NWP models. Techniques to homogenize observation data with the purpose of generating long observational time series exist, but the homogenization of NWP-model data is a more challenging problem.

To achieve ideal training data sets containing representations of many possible environmental conditions (including extremes) many past years of forecast data from ensemble prediction systems with a homogeneous design are required. The ideal solution would be to retrospectively generate past forecasts with the latest model version, restarting from previous model initial data, resulting in *reforecasts* (or *hindcasts*), all other features being equal. For example, reforecasts are provided by global prediction centers such as the U.S. National Weather Service or the European Center for Medium-range Weather Forecast (Hamill et al., 2013). As discussed by Hamill (2018), this procedure is computationally very demanding and may impact many other aspects of the centers' activities. The added skill coming from additional training data for postprocessing methods thus needs to be weighed against the loss of computational resources for EPS model improvements, for example by running the model at a higher resolution.

Many postprocessing schemes using time-adaptive training sets (e.g. sliding windows) have been developed to cope with that problem. They typically have to rely on past forecast cases from only the current and possibly a few past seasons (Wilson and Vallée, 2001). Even though these methods adapt to new data being progressively added to the training data, eventually nullifying the need for a statistical model adaptation, temporary degradation of the postprocessing following model changes can often be observed (Lang et al., 2020), and may impact the overall performance of the statistical model. Therefore, the development of methods to better incorporate model changes into postprocessing models has recently received substantial research interest. For instance, a new method based on the use of the trajectories of the linearized NWP model to possibly reduce the cost of homogenizing the past-forecasts database following a model change was recently proposed (Demaeyer and Vannitsem, 2020). The effects of model changes will typically vary substantially according to the weather variable and context. For example, it may be essential for modeling rare events or small-scale phenomena to use a longer training period containing model changes rather than relying on a shorter period where reforecasts are available (Hess, 2020).

The use of ML methods for postprocessing presents new opportunities to account for model changes during training (as well as suffering themselves from model changes). For example, binary indicators of changes of the NWP model versions could serve as additional predictors in the postprocessing model (Hess, 2020). Further, information on the model change could be incorporated with linearized forecast models, possibly allowing ML algorithms to benefit from the trajectories of the linearized model (and the information of the model change encoded inside) as training data.

## 5. Blending multiple forecasts

Conceptually, the blending of multiple forecast sources aims to create a single improved forecast (either by gaining continuity in transition or greater combined skill), based on the assumption that the inputs are samples from the same Probability Density Function (PDF). In operational centers, the blending of multiple forecast sources has tended to be deterministic and focused on two forecasting concerns; firstly, optimizing the inclusion of a radar extrapolation nowcast (e.g. Golding, 1998; Bailey *et al.*, 2014) and secondly, incorporating multiple forecast sources (Woodcock and Engel, 2005; Engel and Ebert, 2012) to improve skill. An example of blending multiple model forecasts is provided in Figure 3.

More recently the focus has included convection-permitting ensembles (Clark *et al.*, 2016, Beck *et al* 2016) and blending methods that are also capable of generating a single seamless forecast across timescales. Blending first requires a re-projection on to a common grid and either temporal interpolation or disaggregation to a common temporal frequency (Woodcock and Engel, 2005). The inputs being blended should represent the same phenomena and that may require some form of calibration or downscaling (Howard and Clark, 2007; Sheridan *et al.*, 2010, 2018; Moseley, 2011) to ensure equivalency between the inputs (e.g. Kober *et al.,* 2012).

There are two fundamental ways of blending, either in the physical forecast space, or in probability space. In physical space, the simplest approach is to compute a lead time dependent weighted average of multiple deterministic forecast sources (e.g. Haiden *et al.*, 2011). For ensemble forecasts, all sources can be treated as equally likely members which simply increase the ensemble size (DelSole *et al.*, 2013; Beck *et al.*, 2016). A deterministic forecast can either be included as an additional member or an ensemble can be generated from the deterministic forecast, e.g. a nowcast extrapolation ensemble (Bowler *et al.*, 2006; Atencia and Zawadzki, 2014). Methods for generating an ensemble forecast from the observed radar truth include STEPS (Seed *et al.*, 2013; Foresti *et al.*, 2016), dynamically changing weight functions (Yu *et al.*, 2015) and the use of an Ensemble Kalman Filter (Nerini *et al.*, 2019). The physical realism of the resulting forecasts is key for providing inputs to applications, such as, hydrological models (Berenguer *et al.*, 2005; Heuvelink *et al.*, 2020).

Probabilistic forecasts are derived by finding the probability of occurrence for a set of thresholds (e.g. T >0°C, T>1°C …). Blending probabilistic forecasts produces a smooth blend, assuming similar properties for the input forecasts, by easily dealing with spatial mismatches between fields (Johnson and Wang, 2012; Kober et al., 2012). The full ensemble distribution is also retained, unlike for physical blending that tends to damp towards the climatological mean, whilst coherent scenarios in spatial structure can be regenerated. Blending probabilistic forecasts, as in Johnson and Wang, (2012), Kober *et al.*, (2012) and Bouttier and Marchal, (2020), typically smooths the grid point forecasts using a neighborhood approach (Theis *et al.*, 2005; Schwartz and Sobash, 2017) to effectively generate more members and create smoother probability fields prior to calibrating using, for example, a reliability-based procedure (Zhu *et al.*, 1996; Flowerdew, 2014).

Optimising a blend of forecast sources relies on the input forecasts having sufficient skill. Computationally expensive blending techniques are unlikely to be beneficial if the input forecasts are very poor. If the input forecast are sufficiently skillful, then a simple blend can add more skill (Johnson and Swinbank 2009, Beck *et al* 2016). Any blending approach involves weighting the inputs. Common ways are simple linear weighting approaches (Golding, 1998, Woodcock and Engel, 2005, Engel and Ebert, 2012) and computing weights by optimizing a verification metric (Atencia et al., 2019; Bouttier and Marchal, 2020, Schaumann et al., 2020). Tuning the weights by lead time, flow dependence (Atencia *et al.*, 2010, 2019), spatial scale (Seed *et al.*, 2013), convective cell identification (Feige et al., 2018, Posada et al., 2019), and regime classification (Kober *et al.*, 2014) can add information, although for the optimal results this must be in conjunction with advanced calibration (Kober *et al.*, 2014). Therefore, an approach in which multiple processing steps are tuned in unison (Bouttier and Marchal, 2020) is appealing for optimising the final result of the calibration and blending. When considering postprocessing multiple variables (Haiden et al., 2011), a strategy that also preserves the relationship between variables is desirable. Overall, the choice of physical or probability blending depends on the desired product. If coherent scenarios for driving hydrological models is the focus (Berenguer *et al.*, 2005; Heuvelink *et al.*, 2020), then a physical space blending is preferable, whilst blending in probability space is efficient, particularly, for operational systems processing a range of meteorological variables.

To summarize, the key challenge of blending multiple forecast sources in physical space is the requirement to maintain realistic physical features within the blended output. This is a particular problem for more discrete fields, such as precipitation, where spatial mismatches between features within the input fields can lead to unphysical behavior in the blended output. Different forecast sources, especially if at very different resolutions, may not represent equivalent phenomena in the same way, which can then make the notion of blending questionable because they don't come from the same PDF. For blending in probability space, a key challenge remains the generation of physical scenarios from the probabilistic output that are coherent in space and time, and then a further difficulty may be retaining appropriate correlations between different variables. The ideal blending technique would be diagnostic-agnostic, applicable across timescales and able to produce physically realistic forecasts, as required for driving downstream models. Whilst various investigations have provided invaluable insights (Bowler et al., 2006; Johnson and Wang, 2012; Kober et al., 2012; Seed et al., 2013), no single universally-applicable and successful blending technique exists to date.

## 6. Challenges in Operational implementation

For postprocessing, a key challenge is the availability of a training dataset within the operational environment with a truth of sufficient quality. Generation of larger training datasets is often prohibitive due to the data volume and timeliness requirements for operational running, and the presence of model upgrades (see section 4). For operational implementation, a smaller training dataset constructed for a rolling training period is therefore typically favored (Allen et al., 2019). Additionally, a choice is required as to whether calibrate gridded or site forecasts. NWP models output gridded forecasts, while high quality observations as a source of truth are usually available at specific site, only. This could pose problems when gridded calibrated outputs are sought for. The choice of calibrating gridded or site forecasts operationally is therefore hugely dependent on the reliable availability of a truth dataset of sufficient quality for calibration. Operational running mandates a compromise between the scientific ideals of the research community and the practicalities of managing a robust operational system.

The transfer from research to operations (R2O) requires considerable effort to configure a research framework into an operational software environment (see Appendix for examples), ensuring proper run-time behavior, achieving acceptance by users, and maintaining the system in the long-term. A gap between research and operations may be perceived, especially if the necessity for the R2O transfer is not sufficiently understood or acknowledged or the R2O process is not included in research projects. Other key challenges for operational implementation are (a) facilitating technical installation (b) ensuring data volumes and timeliness; (c) producing outputs that are attractive for users, and (d) maintaining the system in the long-term.

### a. Facilitating Technical Installation

The inclusion of statistical post processing into an operational forecast chain can add a great deal of complexity. Initial operational implementation requires several key elements listed here:

- Compliance with a technical production environment that is readily available at the NMS (e.g. coding language, data formats)
- Having suitable quality control and verification of the input and output data, especially in automated production to avoid unphysical customer products
- Ensuring robustness of the production pipelines
- Managing the structuring and archiving of large volumes of forecast and observational data for training

Further developments require communication with users, testing, and potentially time for adaptation by downstream users. In order to ensure a smooth flow of data for customers, separate development and operational production chains are necessary. During the development phase, performance would ideally be demonstrated in a trial using historic data, of sufficient length to probe multiple seasons and reliably measure performance for rarer events. The performance of the operational system needs to be similarly monitored to confirm the implementation provides similar performance to the trial, and to identify problems as they arise, record the quality of the data actually sent to customers, and build a further data archive that can be used to inform future development. New trials are needed when NWP models are changed.

### b. Data volumes and timeliness

Operational weather forecasts (using both NWP and postprocessing) are constrained to run in a short period of time, since forecasts are useless if produced too late.

Traditionally, postprocessing is performed at station locations. This represents up to several thousand points, depending on country size and post-processed element (many more for temperature than wind speed for example). But in many countries, there is a now a requirement for post-processed fields on the model grid at kilometric scale, which may represent several million gridpoints. The timeliness problem is compounded by the introduction of more advanced machine learning techniques, since storing and loading into memory hundreds of regression trees for instance, instead of few coefficients is much more computationally expensive.

Taillardat and Mestre (2020) describe how stored elements may reach several thousand Gigabytes of regression trees, for a production at several million points. These data can be handled by HPC through massively parallel computation capabilities, and extremely fast I/Os. Nevertheless, it takes a great deal of optimization to ensure proper use of HPC in real time.

### c. Attractiveness for Users

Attractiveness for users is essential for maintaining confidence in an operational capability. Users are looking for benefits from postprocessing that may vary. The exact definition of a "good" forecast and the most important technical characteristics depend on the user group and the communication channel used (e.g. visualization for forecasters, open data for external specialized users, smartphone app for the public). The main factors are:

- Relevance and usability of the postprocessed output, i.e. choice of provided variables (including multivariate combinations), regions, output resolution (time/space) and lead times. In addition, there should be no inconsistencies, missing data, jumpiness or unphysical output. There should also be a consideration of pragmatic requirements for which skill is not the only relevant factor and good visualization.
- Explanation of the output. This is usually realized by documentation, newsletters, FAQ tables, additional first level support, training courses, focus groups, etc.
- An interactive approach in which feedback from users includes responses from developers. A thoughtful agreement on the scope of the postprocessing and quality criteria should be made with the users. For instance: on the need for information at observed sites only, or also at unobserved ones; a seamless appearance of the forecasts at different lead times; realistic variability in space and time; suitable verification scores for optimization; and coherence between deterministic and probabilistic forecasts.

### d. Maintaining the System in the long term

Long term maintenance also poses challenges. A key practical consideration is keeping the postprocessing system as portable as possible, allowing for ease of changes in operational infrastructure, for instance by using Docker. Another issue is maintaining the necessary in-house knowledge. Scientific and technical knowledge about the postprocessing system should be easily transferable from one person to another. This may be facilitated by good software development practices (i.e. readability, version management), and up-to-date technical and scientific documentation. Finally, adaptation to upcoming changes in input data and in the technical production environment at the NMS is vital otherwise a postprocessing system can quickly become outdated. It means avoiding having too many different systems that all have overheads.

## 7. Future prospects on postprocessing

The operational implementation of statistical postprocessing techniques for multiple forecasts at very high spatial resolution and for time ranges from minutes to seasons is a challenge faced by all meteorological centers in the world. Despite improvements in NWP-model forecasts, the demand for local high-quality forecasts is increasing, and therefore statistical postprocessing should be an integral part of the operational chain, and high priority is given to ongoing developments. This paper outlines the current advances and challenges faced by the meteorological community in this area.

A first key challenge is to effectively implement techniques that already provide, or have the potential to provide, important improvements to the forecasts (e.g. Hemri et al, 2014). There are currently considerable efforts in that direction at the meteorological services, usually in close collaboration with the academic research as illustrated in Sections 2, 3 and 4. In view of the considerable increase of model resolution and processes involved, new tools should also be envisaged, such as the use of Machine Learning or blending approaches. Although these techniques have great potential for further improving forecasts, they should not be considered a panacea that will solve all the forecasting problems. No statistical method can go beyond the information that is contained within the input datasets, in particular if forecast trajectories are poor because of sensitivity to initial conditions. In order to really be able to benchmark the value of new methods, a common platform on which the different techniques can be compared on a set of appropriately chosen meteorological forecasts is highly desirable.

A second key challenge is the allocation of appropriate computational and staff resources. It may be simply impossible to implement some techniques. For instance, training and inference with comprehensive machine learning approaches can become a significant endeavor. Customized hardware to run NWP systems in use at most NMS and research centers may not be adequate for NWP postprocessing that is attempting to implement big data methods. Appropriate hardware and software should be allocated, such as the GPUs that provide opportunities for massively parallel implementation. Well trained staff is also required for both the research, development and implementation.

A third challenge of NMS is to choose the appropriate techniques for their own or customers' purposes, as discussed in Section 6. There are a large number of possible approaches, and those chosen should be compatible with end-user purposes. As discussed in detail in the WMO publication on postprocessing (WMO, 2020, in preparation), simple approaches like bias corrections can be easily implemented even with very limited IT resources or manpower provided suitable observations are available, but others are much more demanding like blending approaches, machine learning or analogs. The trade-off between complexity and requirements should be assessed beforehand.

A fourth challenge, that was not touched upon in the present review, is the use of new datasets provided by crowdsourcing. These are proliferating rapidly and considerable efforts are now going into determining how to use them in NWP. Statistical postprocessing can of course benefit from these new sources of data for training the algorithms but also to allow more independent verification of the quality of the corrections. Crowdsourcing also introduces some new problems that need to be solved. For instance, when constructing a common calibration model over a large domain the observations may become highly variable in coverage. Furthermore, unstable availability may introduce unrealistic jumpiness for short lead times/nowcasting when updating calibrated forecasts.

A fifth challenge is to develop statistical postprocessing or blending approaches that could provide forecasts at any location, and not only at specific grid points or specific station sites. One option would be to include the location and variables derived from elevation and surface type data as features in regression models in order to make them valid for whole domains. Another strategy would be to remove spatial heterogeneity in the training data before the statistical modeling and then retransform in the end (Dabernig et al., 2017). A third possibility is to apply spatial interpolation methods to well calibrated forecasts at stations (Scheuerer and König, 2014; Taillardat and Mestre, 2020)

Several centers are regularly running reforecasts, but many questions arise about the number of ensemble members or number of past reforecasts are needed for a specific method. The feedback from the postprocessing experts is an important challenge as it requires communication between the different meteorological centers, and to find compromise between the main needs and the forecasting capabilities.

**Appendix: Available tools**

The development of modern postprocessing methods is driven and facilitated by the availability of efficient, cross-platform open source software libraries such as ranger (Wright and Ziegler, 2017) for random forests, or Keras (Chollet et al., 2015) and Tensorflow (Abadi et al., 2016) for neural networks. These modern tools complement available implementations of postprocessing schemes from research software packages such as the R packages crch (Messner et al., 2016), CSTools (Van Schaeybroeck and Vannitsem, 2015) or ensembleBMA (Fraley et al., 2011), all available on https://CRAN.R-project.org/. Operational routines at NMS are also available such as the UK Met Office's IMPROVER library (https://github.com/metoppv/improver) or the Finnish Meteorological Institute Himan tool (https://github.com/fmidev/himan). For a recent overview of computer software for postprocessing and usage examples see Messner (2018).

## List of Figures

**Figure 1**: Overview of probabilistic postprocessing methods, with parametric approaches shown in blue and nonparametric approaches shown in orange. The horizontal axis refers to requirements in terms of training data, tuning parameter choices, available software implementations, and model size. The vertical axis characterizes model flexibility and adaptability in terms of in terms of the applicability to different output variables, and the complexity of representable relationships between inputs (including multiple predictors) and output.

**Figure 2**: Illustration of two methods to preserve correlations of high-dimensional corrected fields for each ensemble member, namely the member-by-member postprocessing (MBMP) and EMOS-ECC. As an example, ensemble temperature forecasts are graphically shown for Frankfurt am Main (Germany, DE), De Bilt (The Netherlands, NL) and Brussels (Belgium, BE). The raw forecasts for each of the indicated ensemble members are correlated among the three locations (upper right panel) and these correlations must also be present after calibration. Using a direct linear transformation, the MBMP method preserves them naturally, while the ECC is necessary to reorder samples from a calibrated probability density function produced using EMOS. From Schefzik (2017).

**Figure 3**: Example of probabilistic blending for a domain centred on the UK. The top panel shows the two NWP models (UKV (1.5 km) and MOGREPS-UK (2.2 km) (Hagelin *et al.*, 2017)) used as inputs to the probabilistic blending on a 2 km grid at lead times of T+8 and T+11, respectively. The central panel shows the output from the probabilistic blending at the specified lead times constructed using a combination of UKV and MOGREPS-UK. The weighting used at each lead time is indicated by the dashed vertical lines in the lower panel, with this panel showing the lead time-dependent contributions to the blended output from the available forecast sources.

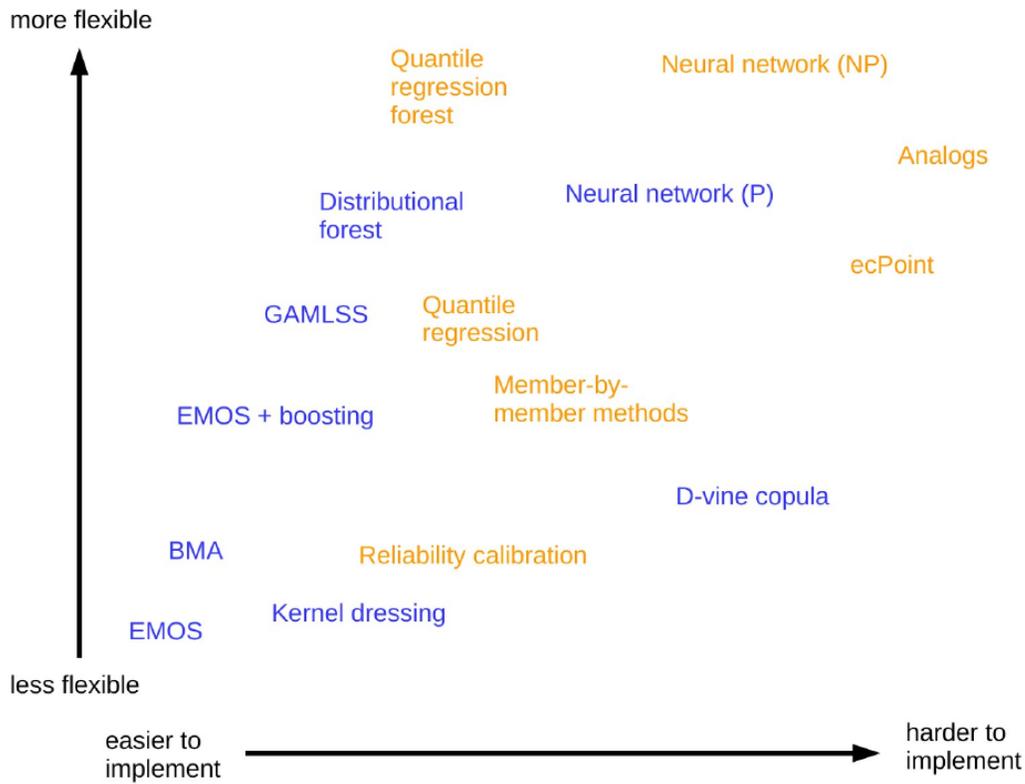

Figure 1: Overview of probabilistic postprocessing methods, with parametric approaches shown in blue and nonparametric approaches shown in orange. The horizontal axis refers to requirements in terms of training data, tuning parameter choices, available software implementations, and model size. The vertical axis characterizes model flexibility and adaptability in terms of in terms of the applicability to different output variables, and the complexity of representable relationships between inputs (including multiple predictors) and output.

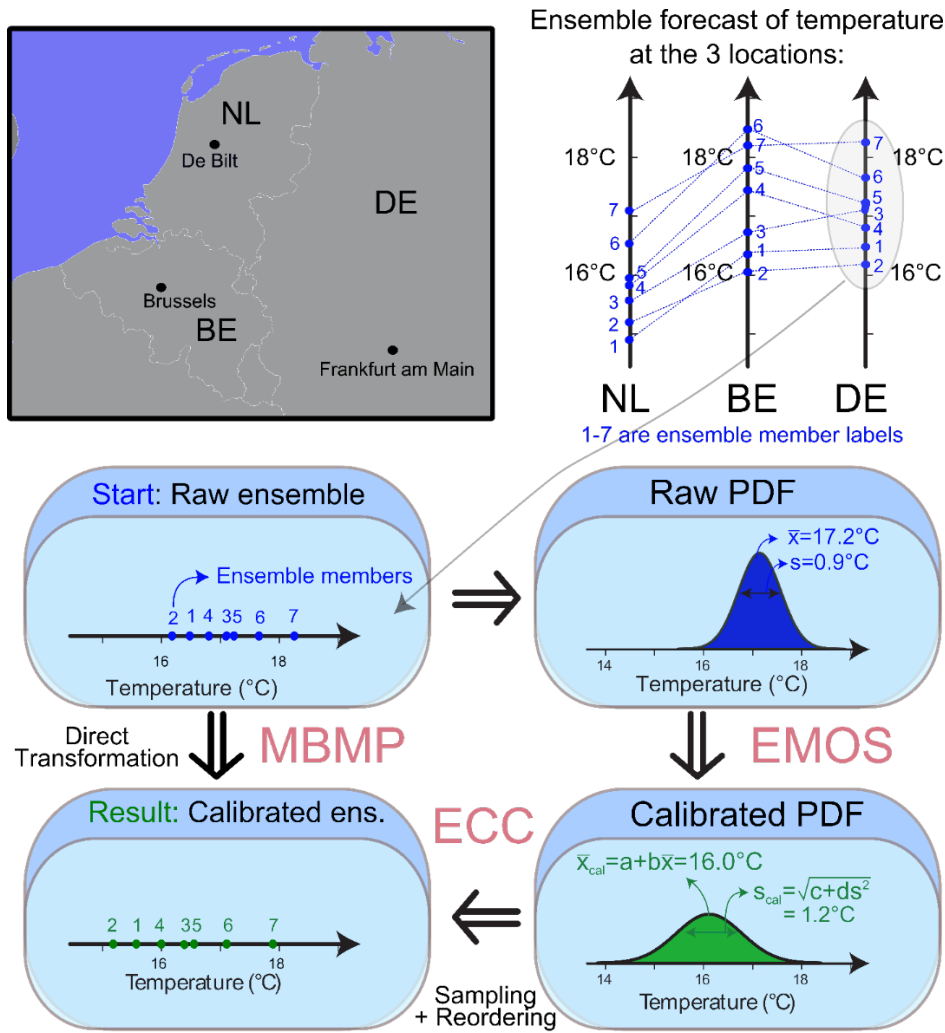

Figure 2: Illustration of two methods to preserve correlations of high-dimensional corrected fields for each ensemble member, namely the member-by-member postprocessing (MBMP) and EMOS-ECC. As an example, ensemble temperature forecasts are graphically shown for Frankfurt am Main (Germany, DE), De Bilt (The Netherlands, NL) and Brussels (Belgium, BE). The raw forecasts for each of the indicated ensemble members are correlated among the three locations (upper right panel) and these correlations must also be present after calibration. Using a direct linear transformation, the MBMP method preserves them naturally, while the ECC is necessary to reorder samples from a calibrated probability density function produced using EMOS. From Schefzik (2017).

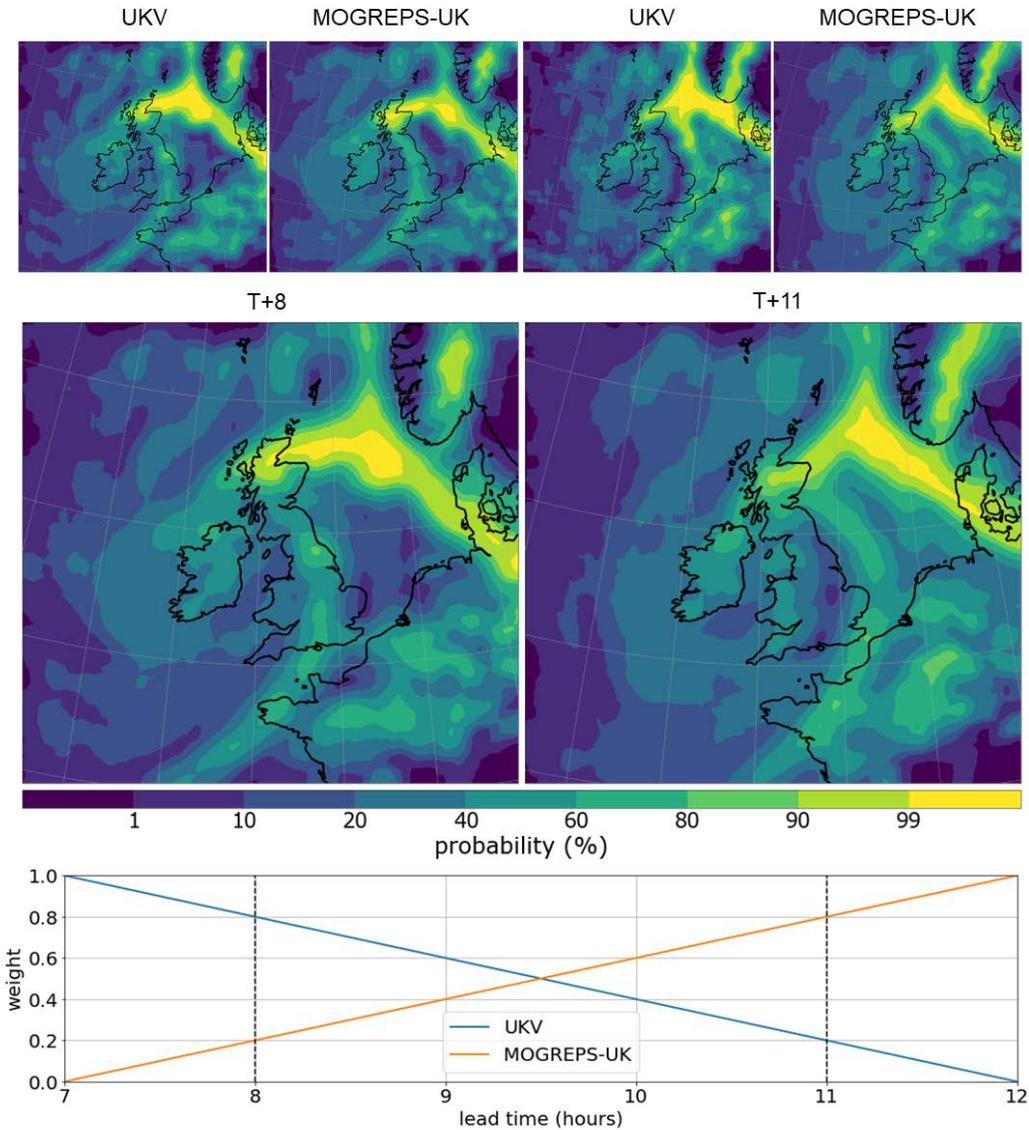

Figure 3: Example of probabilistic blending for a domain centred on the UK. The top panel shows the two NWP models (UKV (1.5 km) and MOGREPS-UK (2.2 km) (Hagelin *et al.*, 2017)) used as inputs to the probabilistic blending on a 2 km grid at lead times of T+8 and T+11, respectively. The central panel shows the output from the probabilistic blending at the specified lead times constructed using a combination of UKV and MOGREPS-UK. The weighting used at each lead time is indicated by the dashed vertical lines in the lower panel, with this panel showing the lead time-dependent contributions to the blended output from the available forecast sources.